# City-Scale Intelligent Systems and Platforms

Klara Nahrstedt  
University of Illinois, Urbana-Champaign

Christos G. Cassandras  
Boston University

Charlie Catlett  
University of Chicago & Argonne National Lab

## 1  Background and Rationale

As of 2014, 54% of the earth's population resides in urban areas, and it is steadily increasing, expecting to reach 66% by 2050. Urban areas range from small cities with tens of thousands of people to megacities with greater than 10 million people. Roughly 12% of the global population today lives in 28 megacities, and at least 40 are projected by 2030. At these scales, the urban infrastructure such as roads, buildings, and utility networks will cover areas as large as New England [China2017]. This steady urbanization and the resulting expansion of infrastructure, combined with renewal of aging urban infrastructure, represent *tens of trillion of dollars in new urban infrastructure investment* over the coming decades [Mynatt2017]. These investments must balance factors including impact on clean air and water, energy and maintenance costs, and the productivity and health of city dwellers. Moreover, cost-effective management and sustainability of these growing urban areas will be one of the most critical challenges to our society, motivating the concept of science- and data-driven urban design, retrofit, and operation—that is, "Smart Cities".

The concept of a 'Smart City' comprises a new generation of innovative services operating within and among key service facets, or *dynamic "systems"* that animate urban infrastructure. Examples include the movement of people, information, and goods; creation and distribution of energy, food, and clean water; effective and equitable provision of education and healthcare; protection and curation of the natural environmental; resources and services enabling economic and social activities; and the provision of public safety and emergency response services [Cassandras2016]. The *technological* underpinnings of these complex dynamic systems, and the facilitation of their interactions, will involve unprecedented numbers of intelligent data sources and actuators whose diversity and volume are equally daunting. These sources and actuators range from administrative systems, to sensor networks, to private and public intelligent urban assets such as vehicles, building control systems, energy and water distribution networks, roadways, and mobile devices—each with unique privacy and security constraints.

Most Smart City projects we see today are relatively simple, predominantly focusing on only one of the service facets listed above. But the ultimate promise of smart cities hinges on the ability to cooperatively optimize these systems, leveraging their interdependencies while mitigating hidden negative "*unintended consequences*." This vision requires platforms and technologies that support integrated analysis of data from many sources, predictive models that capture the behavior of individual urban systems as well as the interdependencies with related systems, automated system controls and interactions, and capabilities for decision-makers at all levels—from individual residents to government officials, and from small businesses to large-scale service providers—to use these integrated data sources and models to inform their own interactions and decisions.

In summary, a Smart City will require intelligent infrastructure that provides: 1) **measurement** of system performance and impact on other systems (including the natural environment); 2) **connectivity** enabling data exchange and cooperative operation among intelligent objects ranging from individual sensors to mobile devices and vehicles, to buildings and roadways; 3) **access** to data in order to support innovation in information provision and decision support technologies and services for individuals, businesses, organizations, and city service providers; 4) **controls** ensuring service integrity, safe autonomous operation, and protection of privacy while empowering individuals and organizations to safely share data as well as physical assets (e.g., vehicle or real estate sharing); and 5) **communication** mechanisms to provide real-time information not only through personal devices but also through integration into physical assets such as street signs [Jaokar2012].

Such an infrastructure represents an evolution of the Internet and the World Wide Web, whose infrastructures are neither the product of independent research nor of natural market forces. Rather, these open, generative infrastructures are the result of a *balance of federal and commercial investment* in basic research and key technologies, and *federally supported communities of academic and commercial participants collaborating toward open infrastructure*. And just as the Internet and Web have fueled (and are in turn today sustained by) entirely new markets and commercial opportunities, an open smart city infrastructure is essential to developing the applications, services, and new markets necessary to support healthy, safe, and sustainable cities in the coming decades.

## 2  Requirements and Challenges of City-Scale Intelligent Systems and Platforms

An integrated, open intelligent infrastructure with the components discussed above—measurement, connectivity, access, controls, and communication—will require building on basic research and emerging technologies and methods, and innovation in areas ranging from material sciences to systems sciences to machine learning. At the heart of these capabilities is data and new software platforms that provide *mobility, security, safety, privacy and processing of massive amounts of information* (so-called "big data"). New capabilities such as sharing of assets or efficient, sustainable design and policy, as well as equitable access to services and opportunities will rely on big data and its use in analysis and modeling. But Smart City infrastructure must also "**close the loop,**" where measurement, connectivity, access, controls, and communication support automated system optimization and responses and improved real-time human decisions and actions [Cassandras2016]. This integration will require a balanced understanding across the range of physical, digital, and social facets of cities embodied in the service facets listed earlier, leveraging the intelligent infrastructure in ways that anticipate and accommodate social, behavioral, and economic forces and effects. Simply put, intelligent infrastructure cannot rely on purely technical and mechanistic views of cities, but on a broader, multi-disciplinary understanding of cities as dynamic interconnected processes rather than merely places.

Increases in population fuel growth in every aspect of infrastructure, and in some cases the value and scale of infrastructure growth rate is super-linear [Bettencourt2007]. Thus city-scale intelligent systems and platforms will need to be optimized for *extensibility and ability to scale* to millions of independent devices, for *high reliability and resilience*, and for *replicability*. Underlying all of these requirements is the importance of **open systems** to support research and development by broad communities using standard hardware and software interfaces [Catlett2017].

The requirements for "Smart City" intelligent systems and platforms suggest a number of basic *challenges*:

### 2.1  Enabling heterogeneous, independent, and widely distributed intelligent devices to interoperate and to exchange data and services.

Development of smart city infrastructure largely parallels challenges faced in the 1980s and early 1990s, as the Internet evolved as an open platform integrating data, voice, video, and specialized applications. The resulting, generative, Internet platform was not the product of industrial organizations, academics, or standards bodies working in a vacuum. Rather, it was catalyzed by federally supported basic research and collaborative contributions by academic, industry, and government researchers who solved fundamental scientific and technical challenges and engaged in their test and evaluation in working infrastructure. The open Internet platform in turn enabled the development of interoperable industry-specific products and services.

In today's Smart City context the "Internet of Things" (IoT) concept is driving many new independent, and non-interoperable platforms, each defining various combinations of mobility, security and safe data collection, embedded (or "edge") computing, and wireless communications technologies to enable massive amounts of data to be analyzed using cloud computing systems.

The open Internet platform is fundamentally a service for transporting packets from one computer to another, underpinning higher-level functionality, beginning with the World Wide Web and ultimately a diversity of interconnected applications. But although the Web is indeed world-wide, with hundreds of millions of independent devices, it is most commonly used to support relatively simple applications involving a handful of interactions, such as loading data from a website. A Smart City infrastructure will require much more **complex data flows**, with applications potentially involving **tens of thousands of heterogeneous devices** that are not only exchanging data but are cooperating, such as to optimize the energy use within a complex of buildings or to re-route traffic around street flooding or collisions. Without the mechanisms to support such orchestration and data integration, each new Smart City application must invent its own unique methods and tools.

What is lacking—and what is necessary to define in the future—is a *common, open, underlying "platform"*, analogous to (but much more complex than) the Internet or Web, allowing applications and services to be developed as modular, extensible, interoperable components. To achieve the level of **interoperation** and innovation in Smart Cities that we have seen in the Internet will require *federal investment in the basic research and development of an analogous open platform* for intelligent infrastructure, tested and evaluated openly through the same inclusive, open, consensus-driven approach that created Internet.

## 2.2    Foundational Support for Privacy, Security, and Reliability

The building blocks of the Internet were largely set in place before it was clear that its growth would be exponential, or that it would become an indispensable utility for nearly every aspect of society. Consequently, the history of the Internet is also characterized by an ongoing struggle to provide security and privacy as the Internet has grown and evolved into critical infrastructure. Most IoT and smart city projects have internalized this lesson, developing elaborate frameworks for privacy and security and thus effectively providing many potential approaches for an open infrastructure. Similarly, many independent IoT or smart city efforts have developed frameworks for reliable system operation and management.

But smart city infrastructure is also physical. Embedding sensing, computing, and communication infrastructure into the built environment requires technology that is as **reliable** as that environment, whether a street or a building or water pipe. Yet technology evolves much more rapidly than the built environment, and thus smart city technologies must also be designed to avoid single points of failure, to provide **self-healing and graceful degradation** in the face of component failures, and to support minimally disruptive upgrades and enhancements. As importantly, the interoperation of physical and digital assets also presents **privacy and security** challenges, requiring consideration of unique physical vulnerabilities [Li2014]). As with privacy, security, and management, today's IoT and smart city efforts have produced many independent, non-interoperable frameworks and platforms.

The landscape for smart city infrastructure today, then, mirrors the 1980s, when there were many non-interoperable communications platforms, including open Internet protocols implemented in NSFNET, specialized federal network protocols such the Department of Energy's ESNET, and proprietary commercial offerings such as from AT&T, DEC, IBM, and MCI. Critical to achieving today's Internet was leadership on the part of the federal government to encourage and support evaluation and selection of a common, open platform, through a process that ultimately engaged all of the platform providers spanning academia, government, and industry in building a common, open Internet.  This leadership is perhaps even more important today for cities than it was for communications networks in the 1980s.

## 2.3    Harnessing New Technologies and Approaches to Sensing and Detection

Smart cities will rely heavily on sensors, whether to measure atmospheric conditions, or air quality, or to detect conditions based on machine learning approaches to analyzing images and sound. Today cities use specialized sensor networks operated by service providers or government agencies, designed and calibrated to provide accurate, actionable data. As with consumer electronics, the price-performance for sensors has steadily improved, to the point that low-cost sensors can be deployed by individuals or embedded in infrastructure such as LED streetlights. Yet the value of traditional sensor networks is not only the technology itself but in its consistent deployment and careful calibration. As sensors proliferate within cities the resulting data will be *massive, noisy and structured* in many different ways. Disruptive technologies such as printed electrochemical [Carter2013] or MEMS-based [Paprotny2013] devices are poised to reduce the **size and cost** of traditionally expensive gas and particle sensing, respectively, further expanding the **diversity and scale of sensor networks**. Consequently a smart city infrastructure must provide common methods for capturing sufficient metadata such that sensor readings can be properly interpreted individually, over time, and in relation to similar measurements from independent systems. [Catlett2017].

Beyond new sensor technologies, **price-performance improvements** in computation coupled with rapidly expanding **machine learning capabilities** present the potential for image and sound processing as a means to detect complex conditions and events, such as "near miss" automobile accidents or the impact of new infrastructure on pedestrian and vehicle flow. Low-cost, high-performance computing technologies also support new approaches involving computation embedded within infrastructure ("*edge computing*"), enabling research into real-time computational control of infrastructure [CMU-ref] and powerful privacy policies, eliminating the need for a centralized server holding potentially sensitive data such as images and sound (as is required to process using central or cloud strategies). An example of such an approach is the NSF-funded Array-of-Things (AoT) [Catlett2017], a partnership between Argonne National Laboratory, the University of Chicago, the City of Chicago, and industry partners. Using Argonne's Waggle platform, designed to provide reliable sensor hosting and edge computing capabilities in remote locations [Beckman2016], AoT is establishing an open platform for rapidly embedding new sensors, communication and computing devices in a major urban areas beginning with Chicago. AoT is also an example of the importance of cities partnering with local universities and national laboratories to harness these new sensing, measurement, and observational technologies, as is also evidenced by the participation of such teams in organizations such as the MetroLab network.

2.4  Resilient Architecture and Balancing Distributed and Central Functions

Many traditional smart city projects rely on data collection from sensors such as cameras, transmitted to a cloud platform where images and other data are continuously and centrally analyzed for decision support. However, this **centralized cloud approach** has several major challenges. First, the **communication costs** can be substantial absent free high-speed networking infrastructure. Second, as noted above, collecting and storing privacy-sensitive information introduces significant **privacy concerns** and challenges regarding public trust and adoption, given that images would most certainly include faces, license plates, and events such as traffic accidents. Third, centralized cloud solutions represent **central points of failure**. Resilient urban infrastructures should not fail if the cloud service is unavailable. Similarly, distributed infrastructure should be resilient enough to fail incrementally ("gracefully"), with self-healing capabilities. **Edge computing** is essential to providing such resilience, but moving from a centralized command-control architecture to a resilient, cooperative, distributed architecture also introduces the need for basic research and development of configuration, **reliable management**, programming, and automated control of distributed systems of autonomous devices.

These distributed systems will involve **tens of thousands of devices,** including **actuators controlling components and functions** ranging from the control of traffic and vehicle flow to the generation and distribution of energy, food, and water. This will require new approaches to providing **real-time adaptation** to natural and/or human-induced disruptions, whether physical or cyber, to ensure **safe and reliable operation**.

2.5  Innovations in Policy, Governance, Privacy, and Ethics

The promise of smart cities cannot be realized without the involvement of its residents, and this requires trust generated by transparency, inclusion, and accountability. **Public engagement, governance and privacy policies** must be an integral part of any city-scale intelligent infrastructure development, test, or deployment. The technical solutions must be accompanies by a set of community-vetted policies and governance structures, reports detailing community input, suggestions, questions and responses and ongoing engagement processes.

3  **Illustrating City-Scale Intelligent Systems and Their Challenges: Transportation**

Among the multitude of functions a city supports, transportation reigns supreme in terms of resource consumption, strain on the environment, and frustration of its citizens. Based on the 2011 Urban Mobility Report, the cost of commuter delays has risen by 260% over the past 25 years and 28% of US primary energy is now used by transportation. Road congestion is responsible for about 20% of fuel consumption in urban areas. In the US, the estimated cumulative cost of traffic congestion by 2030 will reach $2.8 trillion – equal roughly to the U.S, annual tax revenue.

In view of these facts, the advent of **Connected Autonomous Vehicles (CAVs)** is a game-changing opportunity for Smart Cities. Intuitively, there are simple arguments for automated vehicles. A computer is simply a better driver than a human: computers can maintain steady cruising speeds which improves fuel efficiency; they are better at processing data whose abundance is now overwhelming humans (e.g., GPS data, weather data, traffic condition data); they can make fast and accurate driving adjustments; they do not get distracted like humans do; and they do not blink and sleep.

There are of course numerous counter-arguments regarding moral and legal issues; there are also significant technical questions and challenges, and these are precisely what motivates the development of a city-scale intelligent infrastructure that can support the deployment of CAVs and their smooth integration into the existing transportation system. However, CAVs face several major challenges:

- **Communication infrastructure**: The key to the success of CAVs is the ability to share massive amounts of data in real-time. The flow of these data needs to include *Vehicle-to-Vehicle (V2V)* as well as *Vehicle-to-Infrastructure (V2I) communication*. In the latter case, for example, vehicles can share information with traffic lights to control their speed and avoid any possible unnecessary stops at intersections. In fact, as confirmed by simulation, it is possible to completely eliminate traffic lights and maintain a continuous flow of vehicles that never stop [Zhangetal2016].
- **Real-time Data Processing**: In an individual CAV, data originate from both local on-board sensors and from other CAVs and the transportation system infrastructure. These data need to be processed mostly locally, but on occasion at remote sites, in real-time. Therefore, it is necessary to design and manage a collection of cloud resources and computational edge facilities which must be strategically distributed within a Smart City to minimize latency.

Another transportation innovation is the advent of **Electric Vehicles (EVs) and Transportation Electrification**. The EPRI (Electric Power Research Institute) and NRDC study confirms that fueling transportation through electricity instead of petroleum reduces significantly emissions of greenhouse gases and other air pollutants [Tonachel2015]. About 60% of carbon pollution from the transportation sector comes from passenger vehicles. If we electrify all of them with renewably generated, zero-carbon electricity by 2050, we will address a huge part of the climate challenge for the transportation. Furthermore, the EPRI-NRDC study shows that air quality impact by 2030 will be significant, contributing to improved public health. Furthermore, EVs have benefits, compared to conventional combustion engine vehicles, such as (1) they are quiet, (2) they offer high torque, and (3) most notably they produce no tailpipe emissions. However, they have their challenges:

- **Limited Range and Long Battery Charging Time**: The current EV technology require extensive network of charging stations to enable EVs to travel long distances as conventional vehicles do. Furthermore, "refueling" an EV takes much longer time than refueling conventional vehicles. To overcome these challenges, battery technologies will need to become much more efficient in terms of duration and charging time. Another major technology advance that will assist in overcoming these challenges will be the *dynamic charging technology*. Dynamic charging technology allows EVs to charge their batteries while moving on the road, and hence make batteries lighter, smaller and EVs overall cheaper.
- **Privacy and Security:** The EVs are tightly coupled with other sensing, communication and computing infrastructures such as smart grid utilities and road infrastructures. Depending on the charging technology, smart grid utilities may gain information about *vehicle/user location*, violating location privacy needs. In case of dynamic charging, it is important that EVs *authenticate* themselves in real-time with charging components, embedded in roads [Li2014], hence a joint location privacy and secure communication is needed.
- **Billing:** If using dynamic wireless charging technologies for EVs, billing for electricity becomes challenging since smart grid utilities need to charge EVs (mobile appliances) for electricity, but often these electric utilities are not in charge of road infrastructures and EVs. Hence, *coordinated and innovated agreements, pricing models, and billing protocols* which are also privacy-preserving, are required and represent a major challenge for transportation electrification.

## 4 Recommendations

City-scale intelligent systems and platforms are major components of the overall integrated and intelligent infrastructure for future cities and megacities and, more broadly, urban settings. Just as today's Internet and World

Wide Web provide an open infrastructure upon which new markets and industries are fueled, an open infrastructure is essential to catalyzing the innovation necessary to achieve the benefits of smart cities. The challenges outlined above, such as those associated with **scale and complexity**, or **privacy and reliability**, will not be solved within independent academic or commercial efforts. Solutions will require collaboration between academics, industry, and various stakeholders within cities. In this sense, *investments made in programs and coordination efforts will be critical* if the U.S. is to reach the goal of healthy and prosperous urban environments [NSF2015]. Equally important is the development of U.S. commercial capabilities to lead in the expansion of global cities and the creation of new cities worldwide. Research investments are beyond the resources of most cities, and early attempts at independent, industry-developed smart city solutions have been inherently costly absent the common platform approach outlined above. **Investments from federal agencies** will be critical to engage academic researchers in collaboration with cities and private companies, and to support major pilot infrastructures as well as open, community-driven efforts to test, refine, and adopt common platforms. Moreover, there are fundamental as well as applied research challenges that are inherently risky, thus difficult for industry to pursue alone The return of this investment will have significantly broader impacts, as the solutions to city-specific problems are just as relevant to other domains that involve rapidly proliferating IoT technologies (e.g., sensor networks, security, energy management, environmental concerns). In view of these considerations, our recommendations are:

- **The availability of data** is at the heart of any successful research related to Smart Cities. Hence, our recommendation is to **facilitate the creation of an open infrastructure to enable sharing of datasets** for researchers to use, test, and compare new ideas and methodologies. Data should become a *public utility,* with consideration for both open data and the controlled authorized use of sensitive data for research purposes. Along the same lines, incentives need to be in place so that both private and public organizations can make data available on a continuing basis.
- We recommend two focus areas for research regarding the development, prototyping, and deployment of city-scale intelligent infrastructures. Cities within developed economies such as U.S. are faced with aging infrastructure, and thus the need for research into **re-purposing, retrofitting, optimization, and better control of existing infrastructure**, including new measurement, data analysis, and autonomous control systems. Concurrently, cities in developing economies are rapidly expanding, and new cities are being built, where there is need for research regarding the **design and deployment of intelligent infrastructures, systems and platforms toward goals such as reduced energy costs, efficient transportation, and equitable accss to healthcare, energy, food, water, and other services**. Both of these areas require research into data integration and analysis techniques and platforms—particularly those harnessing machine learning—as well as measurement methods and technologies necessary for infrastructure and investment evaluation, and multi-scale, coupled computational modeling capabilities to support optimized designs through simulation.
- The availability of data motivates **data-driven approaches** for resource allocation at all levels of the Smart City infrastructure. Hence, research is needed to develop **data-driven resource allocation mechanisms, including control and audit systems to support privacy and security**—enabling data-driven approaches that go beyond open data, harnessing sensitive data essential to fully understanding or modeling urban systems.
- New research is needed in **machine learning and data analytics** emphasizing dynamic system aspects since city-scale systems and platforms are dynamic complex systems. We need to move beyond static optimization and towards **effective and tractable dynamic optimization under uncertainties**. This focus is a direct consequence of the fact that sensing is now ubiquitous, and we need to shift towards understanding, estimating and inferring higher level knowledge from the raw data in order to optimize and control city-scale systems.
- We recommend research in **distributed/massively parallel algorithmic and information architectures** which can result in computation and information exchange tractability. This includes information architectures which enable processing of **multiscale information from massive data**. For example, in the transportation domain, these data originate from travelers, vehicles, and the transportation infrastructure. What was traditionally modeled as "uncertainty" (noise or disturbance) is rapidly becoming additional input or extra state information with the potential to provide additional context and account for changes that cannot be predicted by deterministic models.
- A common approach to embedded sensing and edge computing platforms needs to inclu research and development of **new sensors, including new materials** for sensing biological materials and pollutants in air and water, **advanced edge computing capabilities** and **resilient embedded hardware systems** such as new computing, sensing or communication devices to operate at low power, and be upgradable, reliable, privacy-preserving, secure, since these systems operate at remote locations and cannot be replaced every 6-12 months (as we assume with our mobile devices).
- Research is necessary to enhance **understanding of interdependencies** in complex dynamic systems since many of the city-scale systems and infrastructure sectors are interdependent and individual sector platforms are constantly changing. We need to understand **coordination mechanisms, protocols, policies and algorithms** to schedule interdependent resources in conflict-free manner to achieve secure and accurate information exchange. For example, as EVs become more ubiquitous, tight coordination will need to occur among sectors such as smart grid utilities, car manufacturers, road infrastructure management and communication companies.
- Research will be needed to explicitly **incorporate human behavior in the operationalization** of Smart Cities. Driverless vehicles (CAVs), for example, could decrease global fuel usage. If, however, CAVs make journeys easier to undertake, fuel usage could increase if population decides to take advantage of this fact. **Understanding paradigm shifts in human behavior** as a result of automation will require new insight into human-technology interactions.

*This material is based upon work supported by the National Science Foundation under Grant No. 1136993. Any opinions, findings, and conclusions or recommendations expressed in this material are those of the authors and do not necessarily reflect the views of the National Science Foundation.*